\documentclass[epj]{svjour}

\usepackage{graphicx}
\usepackage{fancyhdr}
\def\makeheadbox{{
 }}
\def\mytitle{My title} 
\def\myauthors{My name}  
\def\mytype{My type of session}
\def\mysession{My session}

\def\mytitle{Determination of the Discovery Potential for Higgs Bosons in MSSM} 
\def\myauthors{Dorian Kcira}    
\def\mytype{Contributed Talk}    
\def\mysession{Colliders - Higgs Phenomenology}

\begin{document}
\title{Determination of the Discovery Potential for Higgs Bosons in MSSM}
\author{Dorian Kcira
\thanks{\emph{Email: dorian.kcira@cern.ch} }%
, on behalf of the ATLAS and CMS Collaborations
}                     
%
%
\institute{Universit\'e catholique de Louvain, Center for Particle Physics and Phenomenology, Louvain-La-Neuve, Belgium}
%
\date{}
\abstract{
The CMS and ATLAS collaborations have performed detailed studies of the discovery potential for
the Higgs boson in MSSM. Different benchmarks scenarios have been studied both for the CP-conserving case
and for the CP-violating one. Results are presented of the discovery potential in the parameter space
of MSSM.
} 
\maketitle
%
\section{The MSSM Higgs Sector}
\label{sec:mssmhiggs}

Supersymmetry (SUSY) at the TeV scale provides an elegant solution to the hierarchy problem through introduction of superpartners to SM particles and cancellation of problematic loop corrections~\cite{bib:supersymmetry}. It allows light Higgs bosons in the context of GUT without fine tuning.
In the Minimal Sypersymmetric (MSSM) extension of the Standard Model (SM) two isospin Higgs boson doublets are introduced in order to preserve suppersymmetry, one of which couples to the down-type fermions and the other one to the up-type. Eight degrees of freedom therefore exits, three of which are absorbed by the Z and W gauge bosons after the electroweak symmetry breaking. This leads to the existence of five elementary Higgs bosons in MSSM. These physical Higgs bosons are the two CP-even neutral scalar particles $h$, $H$, one CP-odd neutral pseudoscalar particle A, and two charged particles $H^\pm$. Four masses of the elementary Higgs bosons are used to describe the MSSM Higgs sector: $M_h$, $M_H$, $M_A$, $M_{H^\pm}$. In addition two parameters are needed which describe the properties of the scalar particles and their interactions with gauge bosons and fermions. The first of them is the mixing angle $\beta$ which is related to the ratio of the two vacuum expectation values (vev) of the Higgs boson doublets: $\tan{\beta} = v_2/v_1$. The second parameter is the mixing angle, $\alpha$, in the neutral CP-even sector.

\begin{figure}
\includegraphics[width=0.40\textwidth,angle=0]{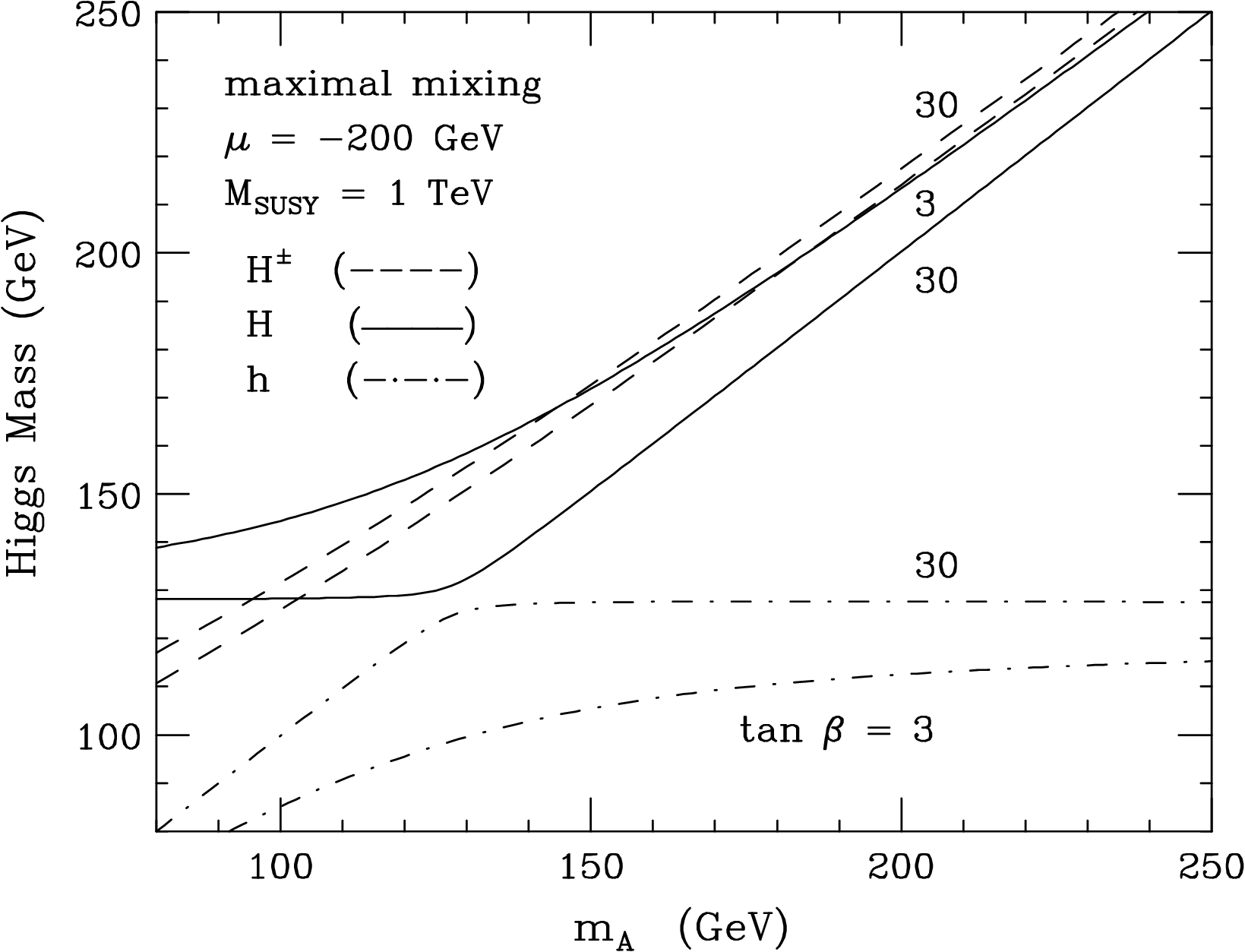}
\caption{The light, heavy and charged CP-even MSSM Higgs boson masses as a function of $M_A$ for $\tan{\beta}=3$ and 30, including radiative corrections. The maximal mixing scenario is used with Higgs boson mass parameter $\mu=-200$~GeV and SUSY mass scale $M_{\rm SUSY}=1$~TeV~\cite{bib:HiggsPheno}.}
\label{fig:radiativecorrections}
\end{figure}

Several relations exist between the MSSM parameters introduced above. At tree-level only two of the parameters are independent and the others can be calculated in terms of them. In the case of CP-conservation the two parameters chosen to describe MSSM are usually $M_A$, $\tan{\beta}$. The following hierarchies hold at tree-level too: $M_h < M_Z$, $M_A < M_H$ and $M_W < M_{H^\pm}$. Such limits would lead to a light scalar Higgs boson in a mass range already excluded by the LEP experiments. This tree-level bound on $M_h$ receives large radiative corrections from SUSY breaking effects in the Yukawa sector of the theory. The leading order corrections are of proportional to $M_t^4$, where $M_t$ is the top quark mass. The upper mass bound for light scalar Higgs boson is reached at large $M_A$ where $h$ becomes SM like and $M_h<135$~GeV. The Higgs boson masses including the radiative corrections are shown in figure~\ref{fig:radiativecorrections}~\cite{bib:HiggsPheno}.

\section{Benchmark Scenarios}
\label{sec:benchmarkscenarios}

Beyond tree-level, the main corrections for Higgs boson masses stem from the $t/\tilde{t}$ and $b/\tilde{b}$ sector (the latter is important for large $\tan{\beta}$). Sub-leading corrections come from all other sectors of MSSM. The Higgs sector phenomenology is thus connected via radiative corrections to the full spectrum on MSSM.  In the unconstrained version of MSSM no particular SUSY breaking mechanism is assumed. A parametrization of all possible soft SUSY breaking terms is used. This leads to more than a hundred parameters (masses, mixing angles, phases) in addition to the SM ones. A detailed scanning of the parameter space becomes impossible. Specific benchmark scenarios are therefore considered~\cite{bib:Benchmarks1,bib:Benchmarks2} for specific points in the parameter space or samples of one- or two-dimensional parameter space. The choice of benchmark scenarios is dependent on the purpose of the investigation: setting conservative exclusion limits, study of typical SUSY experimental signatures, testing of pathological regions of parameter space (worst-case scenarios), etc.


In the CP-Conserving (CPC) scenarios all parameters are real and the neutral mass eigenstates are equal to the CP eigenstates. Only the CP-even bosons couple to weak gauge bosons. The scenarios considered are as follows.
1. {\bf $M_h^{\rm max}$ scenario:} yields the largest $M_h$ value. It was designed to obtain conservative exclusion limits on $\tan{\beta}$ and was used at LEP.
2. {\bf No-mixing scenario:} is similar to the {$M_h^{\rm max}$ one but with vanishing mixing in the $\tilde{t}$ sector and higher SUSY mass to avoid the LEP Higgs bounds. It yields the smaller $M_h$ values (below 116~GeV).
3. {\bf Gluophobic scenario:} effective Higgs-gluon coupling suppressed over large area of the $\tan{\beta}$-$M_A$ plane. Main production channel $gg\rightarrow h$ suppressed, $M_h<119$~GeV.
4. {\bf Small $\alpha$ scenario:} Higgs boson branching ratio into $b\bar{b}$ and $\tau^+\tau^-$ strongly suppressed for large $\tan{\beta}$ and moderately large $M_A$, $M_h<123$~GeV.

\begin{figure}
\includegraphics[width=0.40\textwidth,angle=0]{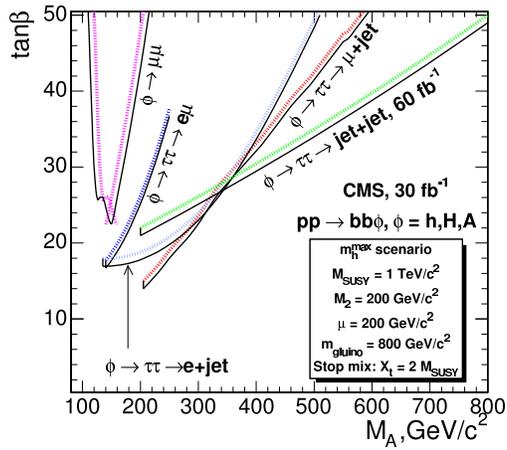}
\caption{$5\sigma$ discovery regions for neutral Higgs bosons $\phi$ produced in association with $b$ quarks in the $M_h^{\rm max}$ scenario with 30~fb$^{-1}$ simulated data for the CMS detector. For more details see text.}
\label{fig:cmsneutralhiggs1}
\end{figure}


In the CP-violating (CPV) scenarios the complex phases related to $A$ and the gluino mass, $M_{\rm gluino}$, constitute additional parameters. A mixing of the CP eigenstates produces the mass eigenstates $H_1$, $H_2$, and $H_3$ in decreasing mass order. No well defined CP can be assigned to them. All neutral mass eigenstates may couple to weak gauge bosons and among each other. {\bf CPX scenario}: the parameters are chosen such that the CP violating effects in the Higgs sector are maximized.  

\section{Results}
\label{sec:results}

The masses of the Higgs bosons, the couplings and the branching ratios for the results from both ATLAS and CMS were calculated with {\sc FeynHiggs}~\cite{bib:feynhiggs} (versions 2.1 and 2.3.2). Most of the signals and backgrounds were generated using {\sc Pythia}~\cite{bib:pythia}.

\subsection{Neutral Higgs Bosons}
\label{subsec:neutralhiggs}

\begin{figure}
\includegraphics[width=0.40\textwidth,angle=0]{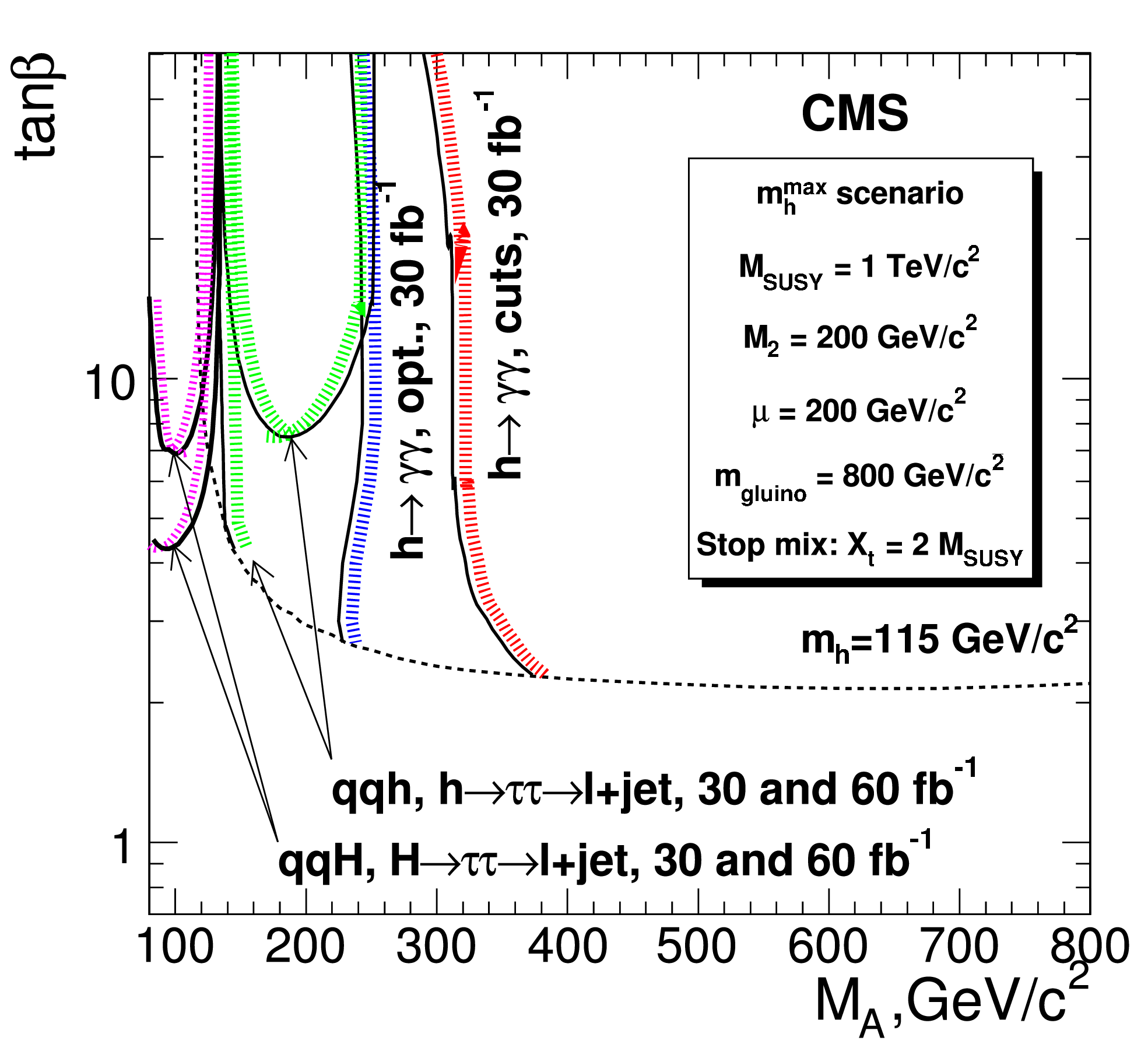}
\caption{$5\sigma$ discovery regions for light and heavy scalar Higgs bosons in the $M_h^{\rm max}$ scenario with 30~fb$^{-1}$ simulated data for the CMS detector. For more details see text.}
\label{fig:cmsneutralhiggs2}
\end{figure}

\begin{figure}
\includegraphics[width=0.40\textwidth,angle=0]{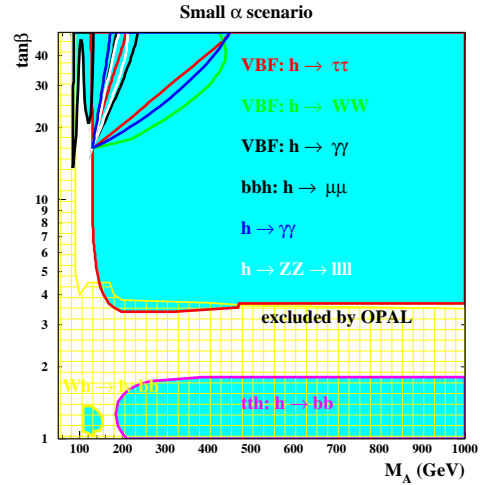}
\caption{Discovery potential for the light CP-even Higgs boson in the CPC small $\alpha$ scenario with  30~fb$^{-1}$ simulated data for the ATLAS detector.}
\label{fig:atlaslightneutralsmallalpha}
\end{figure}

Neutral Higgs boson production was studied in the $M_h^{\rm max}$ scenario with 30~fb$^{-1}$ simulated data of the CMS detector. The results of the $5\sigma$ discovery potential are presented in figure~\ref{fig:cmsneutralhiggs1} for neutral Higgs boson $\phi$ ($\phi =h, H, A$) produced in association with $b$ quarks $pp \rightarrow b\bar{b}\phi$ with  $\phi\rightarrow \mu\mu$ and $\phi\rightarrow\tau\tau$ decay modes. The supersymmetric Higgs boson mass parameter was set to $\mu=200$~GeV. The discovery region for the light, neutral Higgs boson $h$ from the inclusive $pp \rightarrow h + X$ production with the $h \rightarrow \gamma \gamma$ decay and for light and heavy scalar Higgs bosons, $h$ and $H$, produced in the vector boson fusion (VBF) $qq \rightarrow qqh(H)$ with $h(H) \rightarrow \tau \tau \rightarrow l + j$ decay is shown in figure~\ref{fig:cmsneutralhiggs2}.
The VBF channel dominates the discovery potential at low luminosity and covers most of the parameter space left over from the LEP exclusion limits.
The discovery potential for the light neutral $h$ for ATLAS 30~fb$^{-1}$ simulated data is shown in figure~\ref{fig:atlaslightneutralsmallalpha} in the small $\alpha$ scenario. The effect of suppressed branching ratios into $\tau$ leptons is visible for $\tan{\beta}>20$ and $20 < M_A < 300$~GeV. This hole in the discovery region is nicely complemented by $h$ decays into gauge bosons from VBF or gluon-gluon fusion.
Differences between the different scenarios are mainly due to the fact that in the same parameter space point the $h$ mass differs, changing the sensitivity of the channels under consideration.

\subsection{Charged Higgs Bosons}
\label{subsec:chargedhiggs}

\begin{figure}
\includegraphics[width=0.40\textwidth,angle=0]{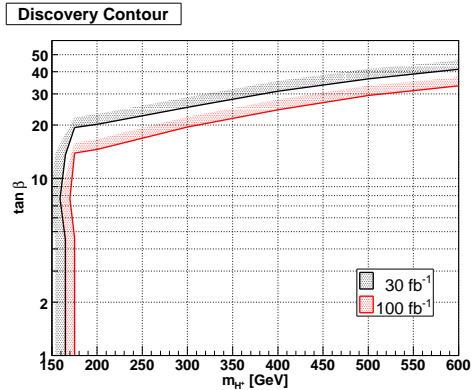}
\caption{Charged Higgs boson discovery contour in the MSSM. The regions above the curves are the part of the parameter space in which a 5$\sigma$-discovery is feasible. Curves for two different integrated luminosities of simulated data with the ATLAS detector are shown.}
\label{fig:atlaschargedhiggs}
\end{figure}

Since no charged scalar Higgs boson is predicted within the SM, search for such a Higgs at LHC is particularly interesting.
At the LHC (and hadron colliders in general) single charged Higgs boson is produced through two main mechanisms:  $gg\rightarrow tbH^\pm$ ($2\rightarrow 3$ process) and $gb\rightarrow tH^\pm$ ($2\rightarrow 2$ process). These two processes are called twin processes~\cite{bib:atlaschargedhiggs} since they correspond to two different approximations describing the same basic process.  The $2\rightarrow 2$ process is dominant at higher masses, $M_{H^\pm}>M_t$, due to resummation of potentially large logarithms by the $b$ quark parton density. In this case the parton shower produces the outgoing $b$ quark.  For Higgs boson masses below the mass of the top, $M_{H^\pm}<M_t$, the $2\rightarrow 3$ process dominates since it incorporates on-shell top quark pair production with subsequent decay to charged Higgs boson. For charged Higgs boson masses range around the mass of the top quark both processes give comparable contributions. Theoretical treatment of this region is complicated and a matching procedure must be applied in order to avoid double counting. The ATLAS collaboration has performed an analysis of simulated data for two different luminosities, 30 and 300~fb$^{-1}$. For the generation of events the MATCHIG~\cite{bib:Matchig} program was used that performs a matching of the above two processes allowing proper handling of the predictions and a consistent tratment of the transition region around the top quark mass. The decay chanels used are $H\rightarrow \tau \nu_\tau,\; t\rightarrow bjj$. Results of the discovery contours obtained for the two luminosities are presented in figure~\ref{fig:atlaschargedhiggs}. An improvement with respect to previous measurements is observed. With 30~fb$^{-1}$ a discovery of charged Higgs boson with mass below 160~GeV is possible for all $\tan{\beta}$, whereas a discovery over the whole mass region is possible for $\tan{\beta}>40$.

\subsection{Overall CPC Discovery Potential}
\label{subsec:distinctionmssmsm}

The overall discovery potential for Higgs boson in the $M_h^{\rm max}$ scenario with 300~fb$^{-1}$ simulated data of the ATLAS detector is shown in figure~\ref{fig:atlasmhmax300}. For all benchmark scenarios in a large part of the MSSM parameter space discovery is possible via several channels, which will allow a determination of parameters of the the MSSM Higgs sector. At least one Higgs boson can be discovered in the whole model parameter space and for a significant part of the parameter space more than one Higgs boson can be observed. This would allow to distinguish between SM and MSSM through direct observation. However in a large area only the observation of the light neutral Higgs boson $h$ is possible. In order to discriminate in this case between a SM or MSSM origin of $h$ the ratio of branching ratios (BR) was calculated for the VBF production $R = {\rm BR}(h \rightarrow \tau \tau) / \rm{BR}(h \rightarrow WW)$. Then a discrimination variable is defined: $\Delta = (R_{\rm MSSM}-R_{\rm SM})/\sigma_{\rm exp}$, where $\sigma_{\rm exp}$ denotes the expected error on $R$. Only statistical uncertainties have been taken into account. The sensitivity of this discrimination variable  is shown in figure~\ref{fig:atlaswedge}. Other studies have been performed for extracting the Higgs boson couplings from the LHC data and testing the sensitivity of the deviations from the SM~\cite{bib:Couplings}. They are based on global fits of the data.

\begin{figure}
\includegraphics[width=0.40\textwidth,angle=0]{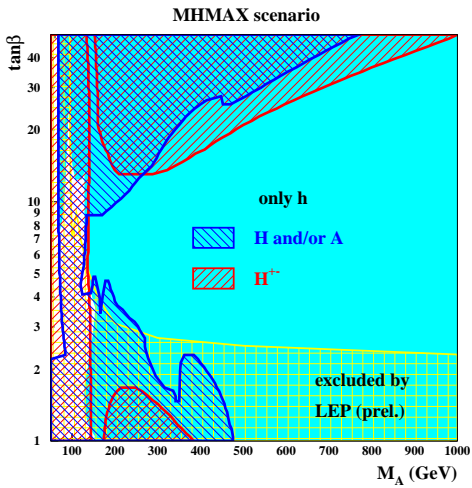}
\caption{Overall discovery potential for Higgs boson in the $M_h^{\rm max}$ scenario with 300~fb$^{-1}$ as calculated with simulated data for the ATLAS detector.}
\label{fig:atlasmhmax300}
\end{figure}

\begin{figure}
\includegraphics[width=0.40\textwidth,angle=0]{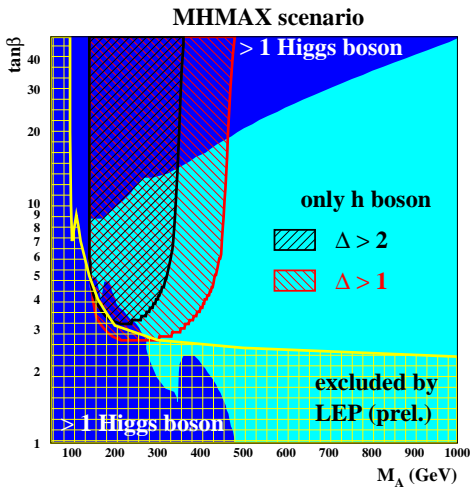}
\caption{Sensitivity for discrimination between SM and MSSM origin of Higgs boson in the $M_h^{\rm max}$ scenario with simulated data for the ATLAS detector. For more details see text.}
\label{fig:atlaswedge}
\end{figure}

\subsection{CPX Scenario}
\label{subsec:cpxresults}

\begin{figure}
\includegraphics[width=0.40\textwidth,angle=0]{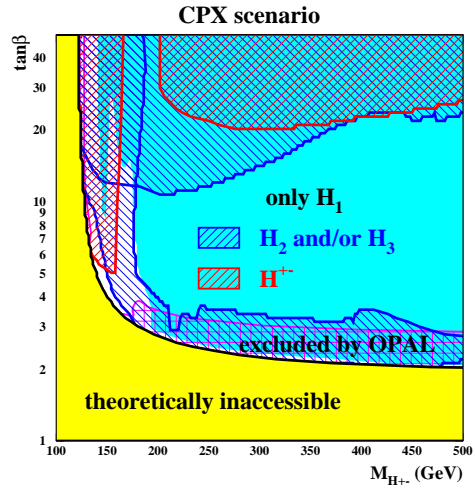}
\caption{Overall discovery potential for Higgs bosons in the CPX scenario with 300 fb$^{-1}$ of simulated data from the ATLAS detector.}
\label{fig:atlascpxallzoom2}
\end{figure}

The overall discovery potential for neutral Higgs bosons in the CPX scenario is shown in figure~\ref{fig:atlascpxallzoom2} for a sample of simulated data of the ATLAS detector with a luminosity of 300~fb$^{-1}$. A coverage similar to the case of CPC scenarios is observed and in most of the parameter space at least one Higgs boson can be observed. It must be noted the that LEP exclusion limits for the CPX scenario are weaker and even low masses $0<M_{H_1}<60$~GeV are not yet excluded~\cite{bib:Opal}. A small region of low $M_{H^\pm}$ and small $\tan{\beta}$ remains where no discovery is possible for the channels and mass ranges investigated in the present studies. The mass of the three CP mass eigenstates $H_1$, $H_2$, $H_3$ is respectively 105, 120 and 140 to 180~GeV. The ATLAS collaboration has started further studies in this particular region of the phase space.

\section{Conclusions}
\label{sec:conclusions}

The discovery potential of ATLAS and CMS Higgs boson searches based on the most recent theoretical calculations and analysis of simulated data has been discussed for four different CPC scenarios and the CP-violating CPX scenario. For the CPC MSSM Higgs boson all the parameter space is potentially covered with already 30~fb$^{-1}$ luminosity of experimental data by discovery of at least one Higgs boson. The coverage observed on all four benchmark scenarios reflects probably most of the MSSM phase space.
In large regions of the phase space only the SM-like MSSM Higgs boson might be observed. Work is ongoing on strategies to distinguish between SM and MSSM origin of the Higgs boson in this case.
For the CPV MSSM almost all parameter space is covered by observation of at least one Higgs boson in the CPX scenario. Studies are underway on small uncovered space ($M_{H_1}<50$~GeV) not excluded by LEP searches.

\section{Acknowledgements}

I would like to thank M. Schumacher, Y. Sirois,\newline M.~Spiropulu and A. Nikitenko for their help in understanding the different aspects of the results presented here. This work was supported by the Belgian Interuniversity Attraction Pole, PAI, P6/11.

\end{document}